# Quantum Computation with Magnetic Clusters

Daniel D. Dorroh[1], Serkay Ölmez[1] and Jian-Ping Wang[2,1*]

[1]*School of Physics and Astronomy, University of Minnesota*

[2]*Department of Electrical and Computer Engineering, University of Minnesota, 200 Union St.,*

*Minneapolis, Minnesota 55455, USA*

*Phone: +1-612-625-9509, fax: +1-612-625-4583, email: jpwang@umn.edu*

# Abstract

We propose a complete, quantitative quantum computing system which satisfies the five DiVincenzo criteria. The model is based on magnetic clusters with uniaxial anisotropy, where standard, two-state qubits are formed utilizing the two lowest-lying states of an anisotropic potential energy. We outline the quantum dynamics required by quantum computing for single qubit structures, and then define a novel measurement scheme in which qubit sates can be measured by sharp changes in current as voltage across the cluster is varied. We then extend the single qubit description to multiple qubit interactions, facilitated specifically by a new entanglement method which propagates the *controlled-NOT* (C-NOT) quantum gate.

# Introduction

In *classical*, digital computation, information is processed and stored in two distinct states which are usually referred to as 0 and 1. A classical computer processes one input at a time to calculate the output. On the other hand, in quantum computing (QC), one can make use of *quantum superpositions* to explore multiple inputs and corresponding outputs simultaneously[1,2]. This feature gives QC important advantages for certain tasks such as searching for an item in an unsorted database. Certain quantum sorting algorithms can complete such a query in $\sqrt{N}$ tries whereas classical algorithms would require on the order of $N$ tries, where $N$ is the total number of items in the database[3]. Additionally, quantum computers can handle problems like the prime-factorization of large numbers[4] which are practically impossible even for supercomputers.

The basic element of a classical computer is a *bit,* which can store 0 or 1. In a quantum computer, the fundamental element is called a *qubit*, which is a physical system with two well-defined states, represented by $|0\rangle$ and $|1\rangle$. The crucial difference between a qubit and a bit is that a qubit can be in a superposition of states, i.e.

$$|\psi\rangle = \alpha|0\rangle + \beta|1\rangle, \tag{1}$$

where $|\psi\rangle$ is the wave function describing the state, and $\alpha$ and $\beta$ are arbitrary coefficients having the property $\alpha^2 + \beta^2 = 1$. Quantum algorithms involve operations on individual qubit states as well as operations that require inter-qubit interactions. A multi-qubit quantum computer must be able to create entangled states by exploiting physical interactions between qubits, which can be realized as Heisenberg type spin-spin coupling, spin-orbital coupling and several other types of interactions. The physical nature of the qubits and the interactions depends on the physical system comprising the quantum computer.

There have been several different proposals for physical implementation of qubits; for instance, some approaches use Josephson Junctions as qubits[5,6,7,8,9,10]. There are also methods using the polarization and propagation direction of photons as the states of the qubits[11]. Nuclear magnetic resonance QC[12,13,14,15] (NMR QC) uses the nuclei of certain molecules to form qubits. A macroscopic number ($10^{19}$) of these molecules are dissolved in a liquid and placed in a strong bias magnetic field (~10 T) along the z-axis. The interaction of the nuclear magnetic dipole with the applied field gives two, well-defined states, spin up and down, represented by $|0\rangle$ and $|1\rangle$ respectively. Furthermore, a small, time-varying magnetic field is applied along the x-axis. The phase and frequency of this magnetic field are precisely adjusted to control the dynamics of individual types of nuclei in the sample to create single-qubit operations. In addition to the interaction with the external field, qubit dynamics is governed by inter-qubit, Heisenberg-type interactions[16]. Making use of the inter-qubit interactions and single-qubit operations, one can implement the *controlled-NOT* (C-NOT) operation, and create entangled states[15], which have been successfully demonstrated in NMR QC. A 7-qubit molecule NMR QC has been implemented and used to factor 15 into its primes[15]. Despite being the most successful QC method, the biggest disadvantage of NMR QC is its lack of *scalability*, since each qubit corresponds to a specific nucleus in the molecule, and each qubit needs a unique Larmor frequency for effective control of its dynamics via the oscillating magnetic field. The need for scalability is the main reason many believe a practical quantum computer should be based on solid state technology for which fabrication techniques are very well developed[17]. In this paper, we are proposing a novel, solid state based QC model using magnetic clusters (MCs) as qubits.

The proposed model satisfies all five DiVincenzo criteria[18], each of which will be discussed below in detail.

# Magnetic Cluster QC

Nanometer size magnetic clusters with high anisotropy have been proposed as qubits[19,20,21]. In this paper, we give a complete discussion of the MC QC including initialization, control and readout schemes. In the proposed MC QC, the qubit consists of a magnetic cluster placed in between two contact terminals, depicted in Fig. 1. Potential candidates for magnetic clusters could be molecular magnets such as $Fe_8$ or $Mn_{12}$, which can be described as quantum systems with large spins at low enough temperatures[22,23,24].

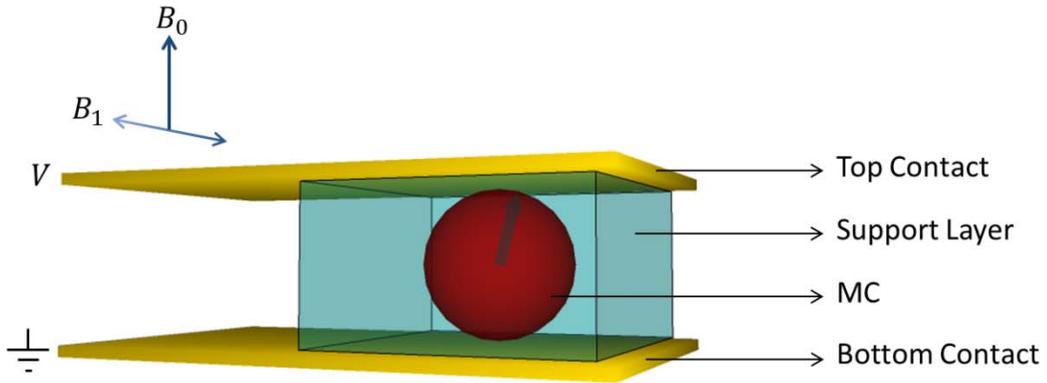

**Figure 1:** *Single qubit structure*. **The magnetic cluster is suspended between two contacts in an insulating matrix. The cluster is chosen to have a small spin ($S < 100$), but a high anisotropy energy. Contacts may be ordinary conducting material (Au). $B_0$ is a fixed, strong magnetic field, whereas $B_1$ oscillates, and the magnitudes of the fields are such that**

**$B_1 \ll B_0$. $B_1$ can be generated using an oscillating current through the top and bottom contacts and a switch (not shown) which allows for single qubit addressing.**

The system is placed in a large magnetic field which is aligned with the anisotropy axis of the MC. Furthermore, a small, oscillating magnetic field is applied in the perpendicular direction to control the dynamics of the qubit. The state of the qubit can be read out by applying a voltage across the terminals, the details of which will be discussed in the Readout section. Before we start discussing the details of the MC based QC, we compare its properties with several available QC approaches, summarized in Table 1.

|  | Kane's QC[25] | NMR QC[12] | SC QC[26] | MC QC |
| --- | --- | --- | --- | --- |
| Physical System | P Nucleus | Nuclei Ensemble | Josephson Junction | Magnetic Cluster ($Fe_8$, $Mn_{12}$) |
| Temperature | Low | Room | Low | Low |
| Scalability | Yes | No | Yes | Yes |
| Challenges | Fabrication & Measurement | Scalability | Decoherence | Decoherence |

Table 1: *Comparison of Different Quantum Information Processing (QIP) Approaches*

The most common challenges amongst the various quantum computing approaches are scalability and difficulty in implementation. In order to perform useful calculations using QC, one needs upwards of hundreds of qubits. This scalability becomes a problem for NMR QC because the experimental resolution required to readout qubits increases with the number of qubits. Similarly, for the case of optics based QC, the number of optical instruments increases exponentially with the number of qubits[11]. Since MC QC inhabits a solid state environment for which fabrication techniques are well established, the scalability problem can potentially be overcome.

## Quantum Description of Magnetic Cluster

The MC, depicted in Fig. 1, has a uniaxial magnetic anisotropy along the axis of the applied magnetic field, $B_0$. When the transverse magnetic field, $B_1$, is turned off, the Hamiltonian that governs the dynamics of this system can be written as:

$$\mathcal{H}_0 = -K\mathbf{S}_z^2 - \gamma\hbar\mathbf{S}_z B_0, \tag{2}$$

where $K$ is the anisotropy energy, $\mathbf{S}$ is the spin operator, $\gamma$ is the gyromagnetic ratio, and $B_0$ is the magnitude of the magnetic field. The eigenstates of the Hamiltonian are labeled by the spin quantum numbers, $S_z = \{-S, -S+1, \cdots, S-1, S\}$, and the corresponding energy levels are shown in Fig. 2.

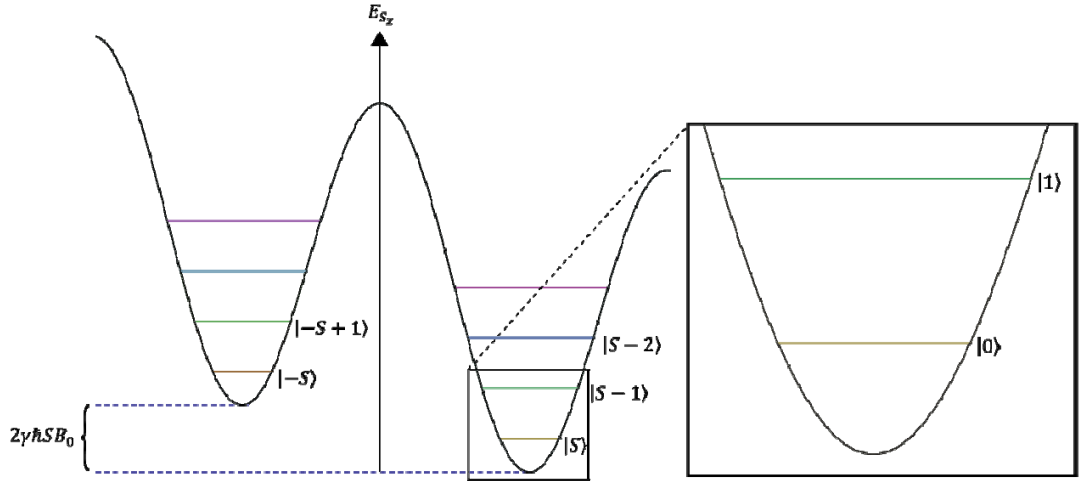

**Figure 2:** *Anisotropic Magnetic Potential Energy and Quantum States.* **The states of the MC in no externally applied magnetic field form a degenerate, double potential well which has minimums when the spin is aligned with the uniaxial anisotropy axis of the MC. The degeneracy is lifted by the applied magnetic field. The two lowest energy states of the MC serve as the qubit states.**

After the initialization, which will be discussed in the *Read out and Initialization* section, the qubit will be in the ground state, $|0\rangle$, which corresponds to $S_z = S$. The first excited state of the qubit is the one corresponding to $S_z = S - 1$, denoted by $|1\rangle$. The energy of each level and the gaps in between are given in Table 2. The energy required to make a transition from the first excited state to the second is $\Delta E \equiv E_2 - E_1 = K(2S - 3) - \gamma \hbar B_0$. Transitions between states are activated by the absorption and emission of phonons[30,31].

| State | Energy | Excitation Energy |
|---|---|---|
| $|0\rangle$ | $E_0 = -KS^2 + \gamma \hbar S B_0$ | $K(2S - 1) - \gamma \hbar B_0$ |
| $|1\rangle$ | $E_1 = -K(S^2 - 2S + 1) + \gamma \hbar (S - 1) B_0$ | $K(2S - 3) - \gamma \hbar B_0$ |
| $|2\rangle$ | $E_2 = -K(S^2 - 4S + 4) + \gamma \hbar (S - 2) B_0$ | $K(2S - 5) - \gamma \hbar B_0$ |

Table 2: *Energies of the First Three States.* Excitation Energy is the energy required to make a transition from a state to the next excited state.

If the temperature is such that $k_B T \ll \Delta E$, where $k_B$ is the Boltzmann constant, the number of phonons with enough energy to make such a transition is exponentially suppressed. Therefore, we can limit the discussion to the subspace spanned by $|0\rangle$ and $|1\rangle$, where $|0\rangle \equiv \begin{pmatrix} 1 \\ 0 \end{pmatrix}$ and $|1\rangle \equiv \begin{pmatrix} 0 \\ 1 \end{pmatrix}$. In this subspace, $\boldsymbol{S}_z^2$ can be represented by a 2x2 matrix, which can be rewritten in terms of $\boldsymbol{S}_z$ as follows:

$$\boldsymbol{S}_z^2 = \begin{pmatrix} S^2 & 0 \\ 0 & (S-1)^2 \end{pmatrix} = (2S - 1)\boldsymbol{S}_z - S(S - 1)\boldsymbol{I}. \tag{3}$$

In this representation, we can rewrite the Hamiltonian as:

$$\boldsymbol{\mathcal{H}}_0 = -K\boldsymbol{S}_z^2 - \gamma \hbar B_0 \boldsymbol{S}_z = -[K(2S - 1) + \gamma \hbar B_0]\boldsymbol{S}_z + \beta \boldsymbol{I} \equiv -\hbar \omega_0 \boldsymbol{S}_z, \tag{4}$$

where $\omega_0 = \frac{K}{\hbar}(2S-1) - \gamma B_0$ and $\beta$ is a constant which is dropped from the Hamiltonian.

Having described the Hamiltonian of the system, we now proceed to discuss how single-qubit operations may be implemented.

### *Single Qubit Operations*

Now, we introduce a time dependent magnetic field of magnitude $B_1$ along the x-axis to generate single-qubit operations. With this magnetic field, the effective Hamiltonian becomes:

$$\mathcal{H} = -\hbar\omega_0 S_z + \hbar\omega_1 \cos(\omega_{mf} t - \phi) S_x, \tag{5}$$

where $\omega_1 = \gamma B_1 \ll \omega_0$, $\omega_{mf}$ is the frequency of the control magnetic field (on the order of GHz), and $\phi$ is the initial phase. Since the Hamiltonian is time dependent, it is convenient to transform the wave function into the so-called *rotating frame*,

$$|\psi_R(t)\rangle \equiv e^{\frac{i}{\hbar}\mathcal{H}_0 t}|\psi(t)\rangle = e^{-i\omega_0 S_z t}|\psi(t)\rangle. \tag{6}$$

Inserting equation (6) into the Schrodinger Equation yields

$$i\hbar \frac{\partial}{\partial t}|\psi_R(t)\rangle = \left(e^{-i\omega_0 S_z t}\mathcal{H} e^{i\omega_0 S_z t} + i\hbar \frac{d}{dt}e^{-i\omega_0 S_z t}e^{i\omega_0 S_z t}\right)|\psi_R(t)\rangle \equiv \mathcal{H}_R|\psi_R(t)\rangle, \tag{7}$$

which shows that the Hamiltonian in the rotating frame is

$$\mathcal{H}_R = \hbar\omega_1 \begin{pmatrix} 0 & e^{-i(\Delta t+\phi)} + e^{-i(\Sigma t-\phi)} \\ e^{i(\Delta t+\phi)} + e^{i(\Sigma t-\phi)} & 0 \end{pmatrix}, \tag{8}$$

where we define $\Sigma \equiv \omega_0 + \omega_{mf}$ and $\Delta \equiv \omega_0 - \omega_{mf}$. Because $\omega_{mf}$ is the frequency of the control field, we can choose it such that $\omega_{mf} = \omega_0$, which gives $\Sigma = 2\omega_{mf}$ and $\Delta = 0$. Furthermore, the terms with $\Sigma$ are rapidly oscillating, and their average becomes zero over the time scale, $1/\omega_1$, which is the time scale for rotations. This approximation is called the *rotating wave approximation*[16]. In this limit $\mathcal{H}_R$ becomes time independent, and reads

$$\mathcal{H}_R = \hbar\omega_1(cos\phi S_x + sin\phi S_y). \qquad (9)$$

The time development operator at $t = \alpha/\omega_1$ can be written as

$$U_R(t = \alpha/\omega_1) = e^{-\frac{i\mathcal{H}_R\alpha}{\hbar\omega_1}} = \begin{cases} e^{-i\alpha S_x} \equiv X(\alpha) \\ e^{-i\alpha S_y} \equiv Y(\alpha), \end{cases} \text{for } \begin{matrix} \phi = 0 \\ \phi = \pi, \end{matrix} \qquad (10)$$

which can be used to generate rotations around the x and y axes. Although equation (10) lacks $S_z$, the rotations around the z-axis in the subspace spanned by $|0\rangle$ and $|1\rangle$ can be generated as a series of rotations around the x and y axes. This follows from the identity

$$Z(\alpha) \equiv X\left(\frac{\pi}{2\sqrt{2s}}\right)Y\left(\frac{\alpha}{\sqrt{2s}}\right)X^\dagger\left(\frac{\pi}{2\sqrt{2s}}\right). \qquad (11)$$

Therefore, we conclude that $|\psi_R(t)\rangle$ can be rotated to any point in the Bloch Sphere, which means any superposition of $|0\rangle$ and $|1\rangle$ can be realized. Now that we have shown single-qubit operations are possible with MC QC, we will turn to the description of how a qubit is initialized and how the final state is read.

# Initialization and Read Out

Qubits of a quantum computer must be initialized before performing an algorithm. After initialization, quantum algorithms require a series of single-qubit and multi-qubit operations. A readout method is required to determine the result of the computation.

To measure the final qubit state, we utilize the *lowest unoccupied molecular orbital* (LUMO)[27,28] of the cluster which can host one or two electrons. A voltage is applied across the qubit to inject electrons into the LUMO to probe the quantum state of the MC. The state of the electron(s) in the LUMO can be represented by $|n, s_z\rangle_e$, where $n$, which can be 0, 1, or 2, is the number of electrons in the state, and $s_z$ is the z-component of the spin of the electron(s). Using this notation, the composite state of the system can be written as $|n, s_z\rangle_e \otimes |m_s\rangle_{MC}$, where $|m_s\rangle_{MC}$ describes the spin state of the MC. The Hamiltonian describing the MC with the addition of electron(s) in the LUMO can be written as

$$\mathcal{H} = -K\mathbf{S}_z^2 - \gamma\hbar B_0 \mathbf{S}_z + (\mathcal{V} - \gamma\hbar B_0 \mathbf{s}_z)\sum_\alpha c_\alpha^\dagger c_\alpha + u\, c_\uparrow^\dagger c_\uparrow c_\downarrow^\dagger c_\downarrow - \frac{J}{2}\sum_{\alpha\beta} c_\alpha^\dagger c_\beta \boldsymbol{\sigma} \cdot \mathbf{S}, \qquad (12)$$

where the operators $c_\alpha^\dagger$ and $c_\alpha$ (indexed with α for spin polarization, ↑ or ↓) are the creation and annihilation operators respectively for electrons in the LUMO, $\mathbf{s}_z$ is the electron spin operator, and $u$ is the electron-electron interaction strength. $\mathcal{V}$ is defined as $\epsilon - eV$, where $\epsilon$ is the energy of the electron in LUMO, and $V$ is the applied voltage. It is important to note that the eigenstates with no electron occupation ($n = 0$) are exactly the eigenstates of the Hamiltonian in equation (2). These states are labeled as $|0\rangle_e \otimes |S\rangle_{MC}$, and $|0\rangle_e \otimes |S-1\rangle_{MC}$ and have energies $E_0 = -KS^2 - \gamma\hbar B_0 S$, and $E_1 = -K(S-1)^2 - \gamma\hbar B_0(S-1)$ respectively. The closest energy eigenstate is $\left|1, \frac{1}{2}\right\rangle_e \otimes |S\rangle_{MC}$. We denote its energy by $E_0^1$, which can be calculated as $E_0^1 = -KS^2 - \gamma\hbar B_0\left(S + \frac{1}{2}\right) + \mathcal{V} - \frac{JS}{2}$. $E_0^1$ changes linearly with the applied voltage since it has one electron in LUMO. All the other states have larger energies either because of the spin polarization of the electron(s), and/or because of smaller $m_s$ values. Fig. 3 shows the energy levels $E_0, E_1$, and $E_0^1$ as a function of $\mathcal{V}$.

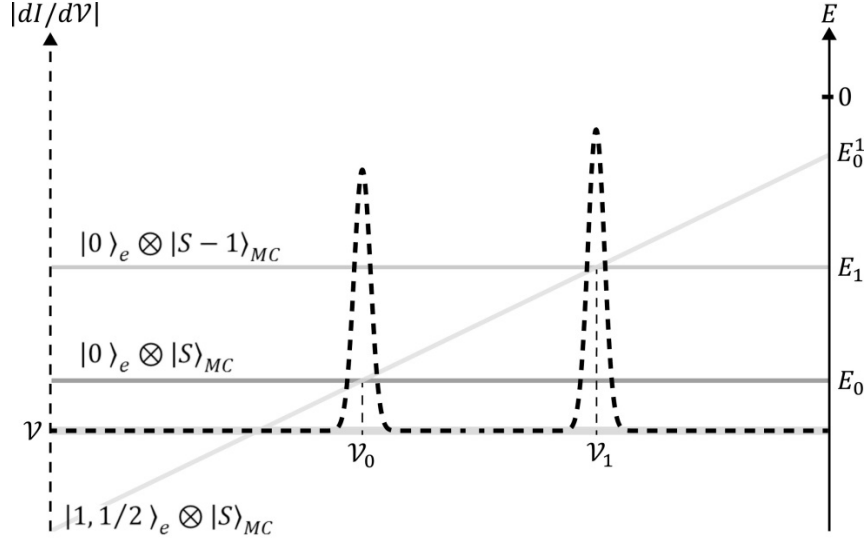

**Figure 3:** *Measurement of Qubit States.* **States are initialized with $V \gg V_1$. After an algorithm is complete, $V$ is lowered. If the qubit state collapses into the $|0\rangle_e \otimes |S\rangle_{MC}$ ($|0\rangle_e \otimes |S-1\rangle_{MC}$) state, a peak will appear at $V_1$ ($V_2$). $|dI/dV|$ peaks correspond to qubit measurements.**

For $V > V_1 \equiv 2K(S-1) + \frac{3}{2}\gamma\hbar B_0 + \frac{1}{2}JS$, the states, $|0\rangle_e \otimes |S\rangle_{MC}$ and $|0\rangle_e \otimes |S-1\rangle_{MC}$, are still the ground and first excited states respectively. In the calculation phase of the algorithm, voltage is applied such that $V \gg V_1$. This will ensure that the LUMO will be unoccupied, and the MC can effectively be described by the Hamiltonian equation (2); therefore, single-qubit operations can be implemented as discussed in the previous section. At the end of the calculation, results are measured in the readout phase.

*Read Out*

To begin a qubit measurement, the voltage is slowly lowered from levels ($V \gg V_1$) maintained during the calculation phase. As seen in Fig. 3, the energy levels of first two states intersect with energy of $\left|1,\frac{1}{2}\right\rangle_e \otimes |S\rangle_{MC}$, therefore transitions from n =0 state to n=1 state are possible. Such a transition is possible by tunneling of an electron from one of the terminals to the MC, which will result in an abrupt change in the current. The value of the voltage at which such a change is observed will give the information on the state of the qubit. Changes in current as $\mathcal{V}$ is varied can be seen in a differential conductance measurement, where peaks in $|dI/d\mathcal{V}|$ correspond to state transitions. A transition peak observed at $\mathcal{V} = \mathcal{V}_1$ would correspond to the collapse of the state to the first excited state, whereas a transition at $\mathcal{V} = \mathcal{V}_0 \equiv \frac{1}{2}\gamma\hbar B_0 + \frac{1}{2}JS$ would correspond to the collapse to the ground state. Electrons traveling from one contact plate to the opposite through the MC cause a current which has been calculated to be on the order of pA[27,28]. Because measuring pA currents through the device may pose challenges, an actual implementation of an MC qubit will likely entail an ensemble of MCs which will allow larger currents. This ensemble approach to realizing MC qubits will be discussed in the *Ensemble MC QC* section. Not only is measurement the key to finalizing a quantum algorithm, but it is also useful in preparing an initial qubit state from which an algorithm can begin.

*Initialization*

The initial state of a quantum computer can be in a superposition of all possible states. To begin a quantum algorithm, the system must be initialized to a known state, which can be done by letting the system relax to its ground state. If the MC qubit is given enough time to relax, the probability of finding it in a state of energy $E$ is approximately

$$P_E \propto e^{-\frac{E}{k_B T}} . \tag{13}$$

The ratio of occupation probabilities for the first excited state and ground state can be estimated as $\frac{P_{E_1}}{P_{E_0}} = e^{-\frac{\Delta E}{k_B T}}$, which quickly converges to zero if $k_B T \ll \Delta E$. Thus, if the system is kept at low enough temperature, qubits will relax into their ground states, which will be used as the initial states of the QC. Furthermore, keeping $\mathcal{V} \gg \mathcal{V}_1$, ensures that the ground state is $|0\rangle_e \otimes |S\rangle_{MC}$.

# 2-Qubit System and Entanglement

Let us consider the 2-qubit system illustrated in Fig. 4. In this setup, the MCs interact with each other by Heisenberg-type coupling.

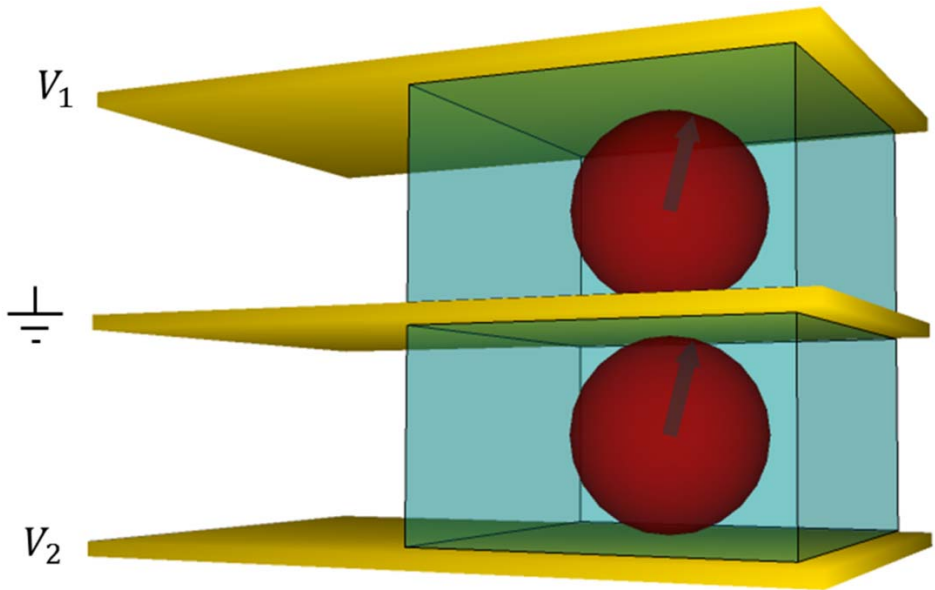

**Figure 4: *Two Qubit Structure*. A two-qubit system can be implemented by aligning two single qubits in a stacked structure.**

The Hamiltonian for this 2-qubit system is given by:

$$\mathcal{H} = -\sum_{i=1}^{2} \hbar\omega_0^i S_z^i + \sum_{i=1}^{2} \hbar\omega_1^i S_x^i \cos(\omega_{mf}^i t - \phi^i) - J\mathbf{S}^1 \cdot \mathbf{S}^2 \quad , \tag{14}$$

where indices label qubits, and $J$ is the interaction strength between the qubits. Following the idea for a single qubit, we transform the wave function into the rotating frame by

$$|\psi_R^{(2)}(t)\rangle \equiv e^{-i\omega_0^1 S_z^1 t} e^{-i\omega_0^2 S_z^2 t} |\psi^{(2)}(t)\rangle, \tag{15}$$

where $\omega_0^1$ and $\omega_0^2$ are the Larmor frequencies of the first and second qubit, and $|\psi^{(2)}(t)\rangle$ and $|\psi_R^{(2)}(t)\rangle$ are the wave functions of the two-qubit system in the lab frame and rotating frame, respectively. If $\frac{\hbar|\omega_0^1 - \omega_0^2|}{J} \gg 1$, we can use the rotating wave approximation to calculate the transformed interaction Hamiltonian, which reads

$$\mathcal{H}_R^{int} = \sum_{i=1}^{2} \hbar\omega_1^i \left(\cos(\phi^i) S_x^i + \sin(\phi^i) S_y^i\right) - J S_z^1 S_z^2. \tag{16}$$

In order to create entanglement, we set $\omega_1^i = 0$, in other words turn off the control magnetic field, and utilize the last term in equation (16). Let us choose the basis states as $\{|00\rangle, |01\rangle, |10\rangle, |11\rangle\}$, and calculate the matrix representation of the time development operator corresponding to the Hamiltonian in equation (16) at $t = \pi\hbar/J$.

$$U_R^{int}(t = \pi\hbar/J) = e^{-i\pi J \mathcal{H}_R^{int}} = \begin{pmatrix} e^{i\pi SS'} & 0 & 0 & 0 \\ 0 & e^{-i\pi(SS'-S)} & 0 & 0 \\ 0 & 0 & e^{-i\pi(SS'-S')} & 0 \\ 0 & 0 & 0 & e^{i\pi(SS'-S-S'+1)} \end{pmatrix} \tag{17}$$

This is the essential operator needed to create the C-NOT operation, which flips the second qubit only if the state of the first qubit is $|1\rangle$. The C-NOT can be realized with the following set of operations

$$C_{2,1} \equiv Y_2\left(\kappa\frac{\pi}{2}\right) U_R^{int}\left(\frac{\hbar\pi}{J}\right) X_2\left(\kappa\frac{\pi}{2}\right) Z_2\left(\frac{\pi}{2}\right) Z_1\left(\frac{\pi}{2}\right) = e^{-\frac{i\pi}{4}} \begin{pmatrix} 1 & 0 & 0 & 0 \\ 0 & 1 & 0 & 0 \\ 0 & 0 & 0 & 1 \\ 0 & 0 & 1 & 0 \end{pmatrix} \quad (18)$$

where $\kappa \equiv \sqrt{\frac{1}{2S}}$, and the overall phase $e^{-\frac{i\pi}{4}}$ can be neglected. The single-qubit operations in equation (18) are defined in equations (10-11), where the subscript denotes the qubit the operators act on. The resultant operation in equation (18) is the C-NOT operation, which can be used to entangle qubits. The description developed for the two-qubit system can easily be generalized to multiple qubits which is done in the next section.

### *Multi-Qubit System*

The Hamiltonian for a multi-qubit system can be written as

$$\mathcal{H} = -\sum_i \hbar\omega_0^i S_z^i + \sum_i \hbar\omega_1^i S_x^i \cos(\omega_{mf}^i t - \phi^i) - \sum_{i<j} J_{ij} \mathbf{S}^i \cdot \mathbf{S}^j ,$$

(19)

where $J_{ij}$ is the coupling strength between the i[th] and j[th] qubit. The coupling strength, $J_{ij}$, may decay quickly with distance separating qubits which makes it difficult to directly entangle two nonadjacent qubits. However, one can still entangle any two qubits by a series of operations on subsequent neighboring qubits. For instance, let us consider a three-qubit system for which there is no direct interaction between qubit 1 and qubit 3 (i.e. $J_{13} \approx 0$). In this case, one can create a C-NOT operation between the first and third qubit as follows:

$$C_{3,1} = C_{3,2}C_{2,1}C_{3,2}C_{2,1}, \tag{20}$$

where the C-NOT operations on the right hand side of the equation are implemented by the couplings between neighboring qubits. This idea can easily be generalized to any multi-qubit system.

# Decoherence

Qubits are physical systems coupled to their environment which can cause the qubits to lose their quantum nature. The decoherence time of a quantum computer is the lifetime of coherent quantum states. In the case of MC QC, the interaction of the MC with the crystal structure and phonons may result in a transition out of the two-state subspace previously defined. In this section, we discuss the dynamics of decoherence, and calculate the lifetime of an MC qubit, focusing on decoherence originating from phonon-assisted transitions out of the qubit's two-state subspace. We turn to references [29,30,31] for the detailed description of the spin-phonon interaction. To begin, we must consider physically what occurs in a single-state excitation facilitated by absorption or emission of a phonon, which is depicted in Fig. 5.

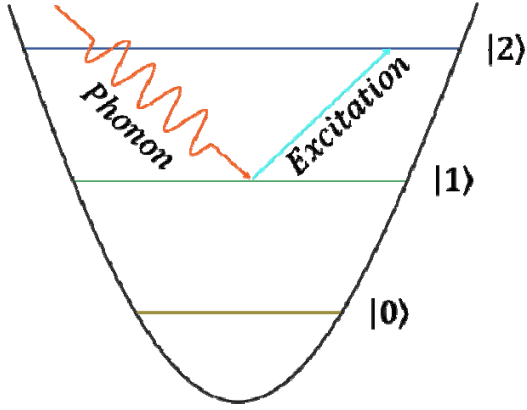

*Figure 5: Phonon Stimulated Excitations*

The states of the system, a few of which are shown in Fig. 5, have different energy and angular momentum values; therefore, any transition between two states requires angular momentum and energy transfer to or from the qubit. Excitations in the crystal structure will carry in (or out) the required angular momentum and energy difference. These excitations can be described as transverse phonons, denoted as $u(r)$. Phonons cause perturbations on the angles of the crystal axes, which can be written as

$$\delta\boldsymbol{\phi}(\boldsymbol{r}) = \tfrac{1}{2}\nabla \times u(\boldsymbol{r}). \tag{21}$$

We can calculate the Spin-Phonon interaction by perturbing the magnetic anisotropy Hamiltonian with the angles $\delta\boldsymbol{\phi}$ as follows:

$$\mathcal{H}_{s-ph} \equiv e^{-i\boldsymbol{S}\cdot\delta\boldsymbol{\phi}}\mathcal{H}_A e^{i\boldsymbol{S}\cdot\delta\boldsymbol{\phi}} - \mathcal{H}_0 \cong (1 - i\boldsymbol{S}\cdot\delta\boldsymbol{\phi})\mathcal{H}_A(1 + i\boldsymbol{S}\cdot\delta\boldsymbol{\phi}) - \mathcal{H}_A = i\delta\boldsymbol{\phi}\cdot[\mathcal{H}_A, \boldsymbol{S}], \tag{22}$$

where high orders in $\delta\boldsymbol{\phi}$ are neglected because $\delta\boldsymbol{\phi} \ll 1$, and the Hamiltonian, $\mathcal{H}_A$, is the anisotropy contribution of the Hamiltonian defined in equation (4).

The decoherence rate corresponding to the transitions from the subspace to outside the subspace can be calculated utilizing equation (22). This is accomplished by calculating the amplitude for the spin-phonon scattering. We define quantum states, $|\Psi_{f,i}\rangle = |\psi_{f,i}\rangle \otimes |\phi_{f,i}\rangle$, where the indices, $f$ and $i$, refer to the final and initial states. $|\psi_{f,i}\rangle$ and $|\phi_{f,i}\rangle$ are the eigenstates of the spin and phonon Hamiltonians respectively. Because two adjacent phonon states differ by one phonon quanta, we define the phonon states, $|\phi_f\rangle = |n_{k,\lambda}\rangle$ and $|\phi_i\rangle = |n_{k,\lambda} + 1\rangle$, where $k$ is the phonon wavevector, and $\lambda = \{t_1, t_2, l\}$ show the transverse and longitudinal polarizations of the phonon. A transition from a state to its adjacent state above is given by the amplitude, $\langle\Psi_f|\mathcal{H}_{s-ph}|\Psi_i\rangle = \Xi \cdot \Phi$, where $\Xi \equiv -i\hbar\omega_{\text{fi}}\langle\psi_f|S|\psi_i\rangle$ is the spin matrix element, $\hbar\omega_{+-}$ is the energy gap between the two states which is listed in Table 2, and $\Phi \equiv \sqrt{\frac{\hbar}{8MN}}\sum_{k,\lambda}\frac{e^{i\boldsymbol{k}\cdot\boldsymbol{r}}}{\sqrt{\omega_{k,\lambda}}}[i\boldsymbol{k}\times\boldsymbol{e}_{k,\lambda}]\sqrt{n_{\omega_{fi}}}$ is the phonon matrix element. Using Fermi's golden rule and the transition amplitude given above, a general transition rate can be defined as

$$\Gamma = \frac{1}{N}\sum_{k,\lambda}\frac{(\boldsymbol{k}\times\boldsymbol{e}_{k,\lambda})^2}{8M\hbar\omega_{k,\lambda}}n_{\omega_{fi}}|\Xi|^2 2\pi\delta(\omega_{k,\lambda} - \omega_{+-}) = \frac{V}{12\pi\hbar}\frac{|\Xi|^2\omega_{fi}^3}{Mv_t^5}n_{\omega_{fi}}, \qquad (23)$$

where $n_{\omega_{fi}} = \frac{1}{e^{\hbar\omega_{fi}^3/k_BT} - 1}$ is the phonon number density, $N$ is the number of cells in the crystal structure, $V$ is the unit cell volume, $M$ is the mass of the cells, and $\omega_{k,\lambda} = v_\lambda k$ is the phonon frequency. $v_\lambda$, the speed of phonons, can be estimated as $\omega_{fi}l_c$, where $l_c$ is the lattice constant. The decoherence time, $\tau \equiv \Gamma^{-1}$, must be long enough such that a large number of single-qubit and multi-qubit operations can be executed before the quantum states decohere. The spin matrix element component for $\Gamma_0$, namely $|\Xi|^2 \propto \hbar^2 S^2 \omega_{\text{fi}}^2$, which can be used to write a proportionality expression for the decoherence time,

$$\tau = \tau(S, T) \propto \left(e^{\frac{\hbar\omega_{+-}}{k_BT}} - 1\right)\frac{Ml_c^2}{S^2}, \qquad (24)$$

is proportional to the moment of inertia of the unit cell. With a small enough $S$ value and properly chosen MC, the decoherence may be made long enough to perform a useful number of operations. The number of operations allowed by a certain decoherence time can be found by comparing the gate operation time, $\tau_g \equiv \sqrt{\frac{1}{2S}\frac{\pi}{\omega_1}} = \sqrt{\frac{1}{2S}\frac{\pi}{\gamma B_1}}$, to the decoherence time in equation (24). Using $\gamma = 0.6 K/T$ where $T$ is Tesla, the application of a magnetic field around $0.01T$ yields a gate operation time, $\tau_g(B_1 = 0.05T) \approx 0.5/K$. We compare several cases of this decoherence time in Fig. 6 below, where we have assumed an anisotropy energy, $K = 0.1 meV$, $l \approx 3 \times 10^{-10} m$, and gap energy $\hbar \omega_{fi} \approx 2KS$.

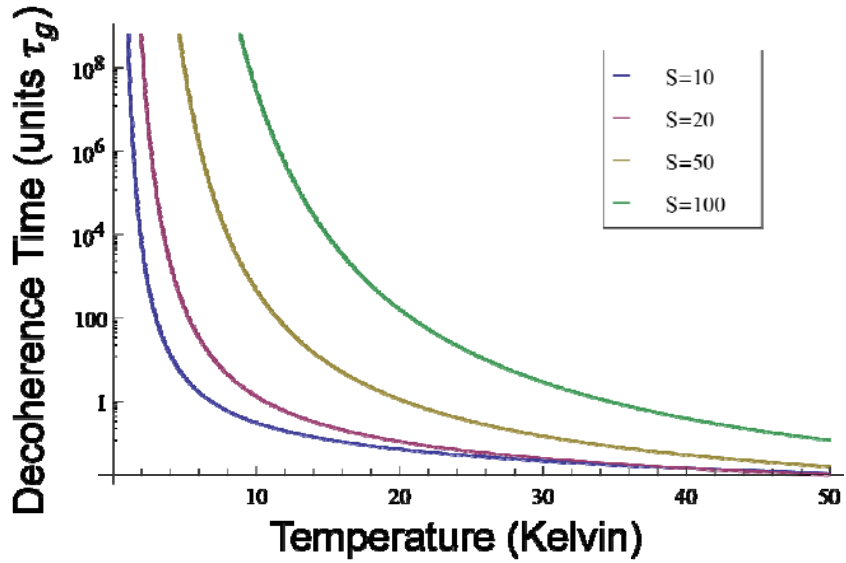

**Figure 6:** *Decoherence.* **Decoherence time is plotted on a log scale in order to characterize the effect of controllable parameters, namely the temperature and the spin of the qubit.**

While Fig. 6 suggests that for MCs with relatively small spin number, $S \approx 10$, decoherence effects are prominent at large temperatures, with sufficient temperature reduction and by choosing an MC with a good spin number, decoherence effects can be successfully suppressed.

# Ensemble MC QC

In order to circumvent challenges of measuring the current through a single MC, an ensemble of MCs can be placed between the contact plates to increase the current. Fig. 7 shows a single-qubit ensemble of MCs, placed in a two dimensional array between the contacts.

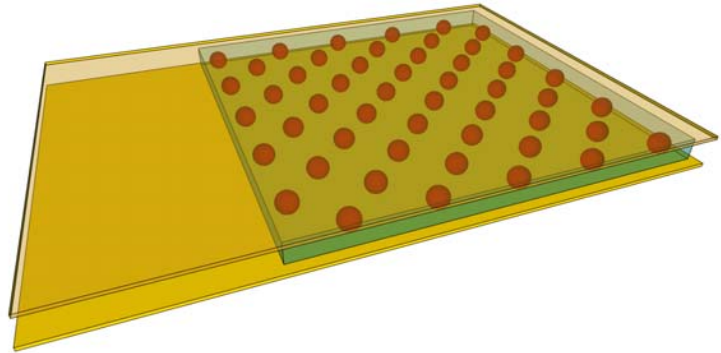

**Figure 7: *Ensemble Qubit.* An array of MCs are placed between two conducting contacts. The distance between each MC is such that interactions between any pair is negligible, which allows us to describe the dynamics of each MC individually.**

The MCs are placed a distance apart such that MC-MC interactions are negligible within the qubit structure. Because these interactions are negligible, the system can be described by the density of states for the ensemble. This approach is in great analogy with NMR QC, for which ensemble computation techniques are well developed[12]. In order to create a second qubit, another layer of MCs are placed directly above the first qubit such that MC grids are precisely aligned. Therefore, each individual MC in the first layer interacts with only one MC in the second layer. This idea can be extended to the case of multiple qubits by building more layers.

# Conclusion

We have provided a theoretical framework for quantum computing using magnetic clusters (MCs) as qubits. With this physical system, we can satisfy all five prerequisites for implementing a quantum computer as follows:

1. The MC qubit is realized in MC spin states which are well-separated energetically, given proper parameters. This energy scales approximately as $2SK - \gamma\hbar B_0$, where $S$ is the spin of the MC, $K$ is the anisotropy energy, and $B_0$ is the applied magnetic field.
2. MC qubits can be initialized via cooling to their ground state due to their asymmetric anisotropy energy structure.
3. A useful number of quantum gate operations can be completed within the decoherence time of the quantum computer.
4. Any two qubits can be entangled via Heisenberg-type interactions and C-NOT propagation. Thus, a Universal set of quantum gate operations can be realized.
5. Qubit-specific measurements can be performed to probe the quantum state using electrons injected into the lowest unoccupied molecular orbital (LUMO).

The MC qubits are scalable because they occupy a solid state environment which provides ease of fabrication and because any two qubits can be entangled despite a lack of direct proximity.

With novel methods for qubit readout, and non-local, qubit-qubit entanglement, we have strengthened the possibility for creating a quantum computer using magnetic components, and have provided an alternative to the currently explored quantum computing proposals.

Technical difficulties associated with the measurement of small currents can potentially be overcome by using an ensemble of non-interacting qubits, as briefly discussed in the *Ensemble MC QC* section. This approach is beyond the scope of the current note, and will be addressed in our following studies.

# Acknowledgement

D. Dorroh and J. P. Wang thank the support from UROP program from University of Minnesota.